\documentclass[aps,twocolumn,preprintnumbers,floatfix,nofootinbib]{revtex4}

\usepackage{graphicx}% Include figure files
\usepackage{dcolumn}% Align table columns on decimal point
\usepackage{bm}% bold math
\usepackage{epsfig}

\usepackage{amsmath}
\usepackage{amssymb}
\usepackage{color}

\def\fun#1#2{\lower3.6pt\vbox{\baselineskip0pt\lineskip.9pt
  \ialign{$\mathsurround=0pt#1\hfil##\hfil$\crcr#2\crcr\sim\crcr}}}

\input epsf

\newcommand{\be}{\begin{equation}}
\newcommand{\ee}{\end{equation}}
\newcommand{\bea}{\begin{eqnarray}}
\newcommand{\eea}{\end{eqnarray}}

\def\pslash{\not{\hbox{\kern-2pt p}}}

\begin{document}

%\vspace*{2cm}

\title{Implications of Gauge Invariance on a Heavy Diphoton Resonance}

%\vspace*{0.2cm}

\author{
%\vspace{0.5cm} 
Ian Low$^{a,b}$ and  Joseph Lykken$^{c}$ }
%\email{ilow@northwestern.edu}
\affiliation{
%\vspace*{.2cm}
$^a$ \mbox{High Energy Physics Division, Argonne National Laboratory, Argonne, IL 60439}\\
$^b$ \mbox{Department of Physics and Astronomy, Northwestern University, Evanston, IL 60208} \\
$^c$  \mbox{Fermi National Accelerator Laboratory, P.O. Box 500, Batavia, IL 60510}
%\vspace*{0.8cm}
}

\begin{abstract}
%\vspace*{0.5cm}
Assuming a heavy electroweak singlet scalar, which couples to the Standard Model gauge bosons only through loop-induced couplings, $SU(2)_L\times U(1)_Y$ gauge invariance imposes interesting patterns on its decays into electroweak gauge bosons, which are dictated by only two free parameters. Therefore experimental measurements on any two of the four possible electroweak channels would determine the remaining two decay channels completely. Furthermore, searches in the $WW/ZZ$ channels probe a complimentary region of parameter space from searches in the $\gamma\gamma/Z\gamma$ channels. We derive a model-independent upper bound on the branching fraction in each decay channel, which for the diphoton channel turns out to be about 61\%.
Including the coupling to gluons, the upper bound on the diphoton branching fraction implies an upper bound on the mass scale of additional colored particles mediating the gluon-fusion production. Using an event rate of about 5 fb for the reported 750 GeV diphoton excess, we find the new colored particle must be lighter than ${\cal O}(1.7\ {\rm TeV})$ and ${\cal O}(2.6\ {\rm TeV})$ for a pure CP-even and a pure CP-odd singlet scalar, respectively. 
\end{abstract}

%\pacs{draft}

\maketitle

\section{Introduction}

The ATLAS and CMS collaborations have recently reported interesting excesses in diphoton searches extracted from LHC Run 2 data of $pp$ collisions with center of mass energy $\sqrt{s}$$=$13 TeV \cite{ATLAS-CONF-2015-081,CMS PAS EXO-15-004}.
The excesses can be interpreted as a new heavy resonance with mass of approximately 750 GeV$/c^2$, and appear to be at least marginally compatible with diphoton searches in LHC Run 1.

With the data in hand, other properties of the putative resonance are very much in doubt. The ATLAS data prefer a large width, while the CMS data prefer a narrow width. The production mechanism at the partonic level is unknown; it could be dominated by gluon fusion, as for the Higgs boson, or it could involve photon fusion \cite{Fichet:2015vvy,Csaki:2015vek}, or quarks \cite{Buttazzo:2015txu,Franceschini:2015kwy}. It is also possible that the decay involves other particles in addition to photons \cite{Cho:2015nxy}.

In this note we consider the possibility that the excesses are the result of a heavy electroweak singlet scalar that couples to Standard Model (SM) gauge bosons via dimension five operators gauge invariant under $SU(2)_L\times U(1)_Y$.
We base our discussion on a CP-even scalar, but similar arguments will apply to a CP-odd scalar unless otherwise noted. For a CP-even scalar we make the additional assumption that we do not need to take into account mixing with the SM Higgs; this is a non-trivial assumtpion since such mixing is expected to be induced by the same new physics that induces the above-mentioned dimension five operators \cite{Berthier:2015vbb}. In other respects our treatment is model independent.

In our simple framework experimental measurements on any two of the four possible electroweak channels would determine the remaining two decay channels completely. Furthermore, searches in the $WW/ZZ$ channels probe a complimentary region of parameter space from searches in the $\gamma\gamma/Z\gamma$ channels. We derive a model-independent upper bound on the branching fraction in each decay channel, which for the diphoton channel turns out to be about 61\%. If we assume that the singlet resonance is produced predominantly via gluon fusion,
we show that the upper bound on the diphoton branching fraction implies an upper bound on the mass scale of additional colored particles mediating the gluon fusion production.

\section{Scalar Couplings with gauge bosons}
\label{sect:section2}

 Couplings of an electroweak singlet scalar with the SM gauge bosons are induced at the one-loop level at leading order. Assuming linearly realized $SU(2)_L\times U(1)_Y$, only three dimension five gauge-invariant operators are generated:
 \bea
 \label{eq:su2u1inv}
&&  \frac{\alpha_{em}}{4\pi s_w^2}  \frac{\kappa_W}{4m_S} S W^a_{\mu\nu} W^{a\,\mu\nu}+ \frac{\alpha_{em}}{4\pi c_w^2}  \frac{\kappa_B}{4m_S} S B_{\mu\nu} B^{\mu\nu}  \nonumber \\ 
 && + \frac{\alpha_s}{4\pi} \,\frac{\kappa_{g}}{4 m_S} SG_{\mu\nu}^a G^{a\, \mu\nu} \; .
 \eea
The  mass eigenstates are defined as
\bea
W^\pm &=& \frac1{\sqrt{2}}(W^1 \mp i W^2)\ , \nonumber \\  
\label{eq:eweigen}
 \left( \begin{array}{c}
              Z\\
               A
             \end{array}\right) &=& 
             \left( \begin{array}{cc}
             c_w & -s_w \\
             s_w & c_w
             \end{array}\right)
 \left( \begin{array}{cc}
             W^3\\
             B 
             \end{array}\right)  ,
\eea
where the sine and cosine of the weak mixing angle are $c_w = {g}/{\sqrt{g^2+g^{\prime 2}}}$ and $s_w = {g^\prime}/{\sqrt{g^2+g^{\prime 2}}}$, respectively. Then the first two operators in Eq.~(\ref{eq:su2u1inv}) become
\bea&&
\hspace{-10pt}
 \frac{\alpha_{em}}{4\pi}  \frac{1}{4m_S}(\kappa_B+\kappa_W) S F_{\mu\nu} F^{\mu\nu}  +\frac{\alpha_{em}}{4\pi s_w^2}  \frac{\kappa_W}{4m_S} S W^+_{\mu\nu} W^{-\mu\nu} \nonumber \\ 
&&\hspace{10pt}
+ \frac{\alpha_{em}}{4\pi}s_w c_w \left(\frac{\kappa_W}{s_w^2} -\frac{\kappa_B}{c_w^2}\right) \frac{1}{4m_S} S F_{\mu\nu} Z^{\mu\nu} \nonumber \\
&&\hspace{10pt}
+  \frac{\alpha_{em}}{4\pi} \left(\kappa_W \frac{c_w^2}{s_w^2}+\kappa_B\frac{s_w^2}{c_w^2}\right) \frac{1}{4m_S} S Z_{\mu\nu} Z^{\mu\nu} \ ,
 \eea
 from which we obtain the following couplings:
\bea
\label{eq:stensor}
\Gamma^{\mu\nu}_{SV_1V_2}&=& \frac{g_{sV_1V_2}}{m_S}  (p_{V_1}\cdot p_{V_2} g^{\mu\nu} -p_{V_1}^\nu p_{V_2}^\mu)  \ , \\
 \label{eq:singscoup}
 % \begin{array}{ll}
g_{Sgg}&=& \kappa_{g} \frac{\alpha_{s}}{4\pi}\ , \\
g_{S\gamma\gamma} &=& \frac{\alpha_{em}}{4\pi} (\kappa_W+\kappa_B)\  ,\\
g_{SZ\gamma}&=& \frac{\alpha_{em}}{4\pi} c_w s_w \left(\frac{\kappa_W}{s_w^2}-\frac{\kappa_B}{c_w^2}\right) \ ,\\
 g_{SZZ}&=& \frac{\alpha_{em}}{4\pi}\left(\kappa_W \frac{c_w^2}{s_w^2} +\kappa_B \frac{s_w^2}{c_w^2}\right)\ , \\
 g_{SWW}& =& \kappa_{W} \frac{\alpha_{em}}{4\pi s_w^2}\ .
\eea
The partial decay widths, in the limit $m_S\gg m_{W/Z}$,  are given by the following expressions \cite{Low:2011gn}:
 \be
 \label{eq:Gv1v2}
 \Gamma_{V_1V_2} =N_g \frac{\delta_V}{64\pi }|g_{SV_1V_2}|^2 m_S \ ,
  \ee
  where $N_g=8$ for digluon final states and $\delta_V=2$ for final states with non-identical particles.

%%%%%%%%%%%%%%%%%%%%%%%%%%%%
\section{Constraints on decays into electroweak gauge bosons}
%%%%%%%%%%%%%%%%%%%%%%%%%%%%%

%%%%%%%%%%%%%%%%%%%%%%%%%%%%%%%%%%%%%%%%%%%%%%%%%%
\begin{figure*}[!t]
  \begin{center}
    \includegraphics[width=0.47\textwidth]{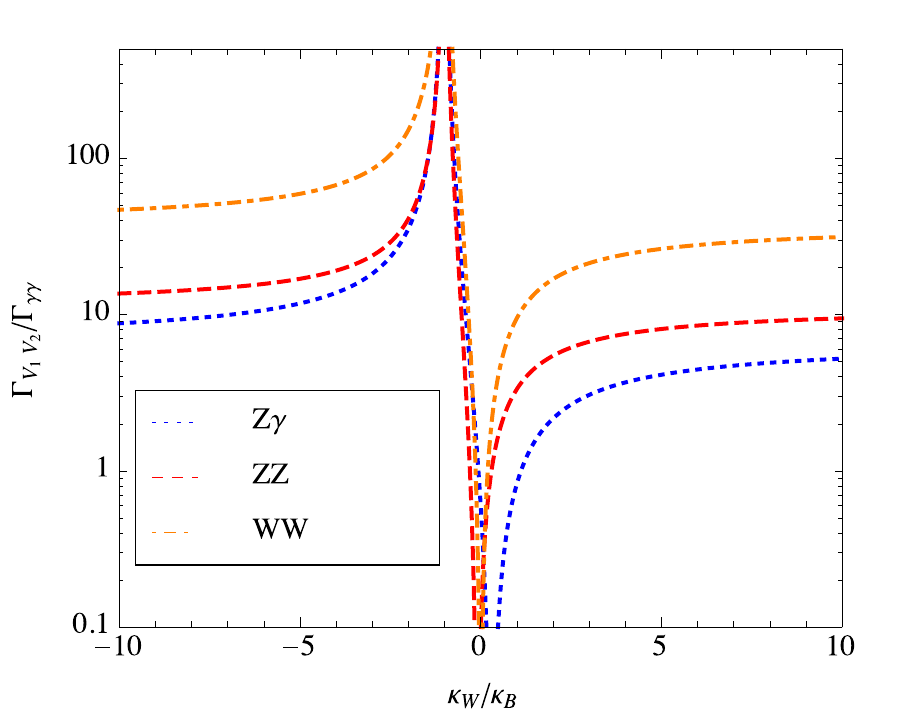}
        \includegraphics[width=0.47\textwidth]{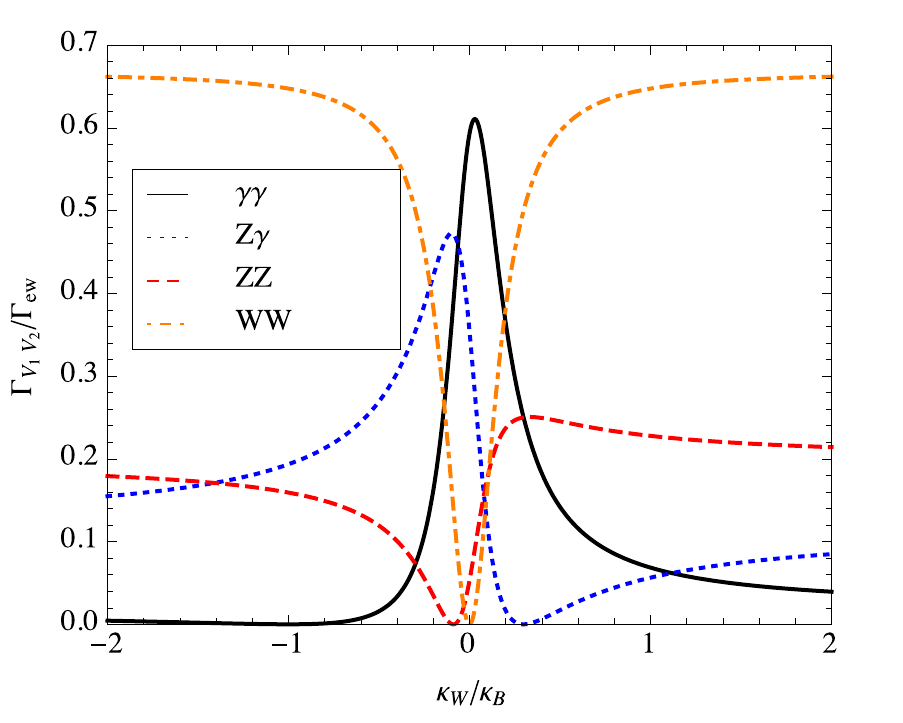}
    \caption{\label{fig:1}{ Left}: Partial widths in ${WW}$, ${ZZ}$ and ${Z\gamma}$, normalized to $\Gamma_{\gamma\gamma}$, as a function of $\kappa_W/\kappa_B$. Experimental input from one channel then determines the partial widths in the other two channels. Right: The upper bound of ${\rm BR}(V_1V_2)$ as a function of $\kappa_W/\kappa_B$, which has a maximum of 61\%, 47\%, 22\% and 67\% for $\gamma\gamma$, $Z\gamma$, $ZZ$ and $WW$ channels, respectively.}
    \end{center}
\end{figure*}
%%%%%%%%%%%%%%%%%%%%%%%%%%%%%%%%%%%%%%%%%%%%%%%%%%

Gauge invariance implies  the partial decay width in the four electroweak channels are determined by only two free parameters: $\kappa_W$ and $\kappa_B$. In other words, if we take any two of the partial widths as input, the other two channels are completely predicted. There are two simple cases: i) $\kappa_W=0$, when the particle running in the loop is an $SU(2)_L$ singlet and only carries the hypercharge, and ii) $\kappa_B=0$, when the loop particle carries no hypercharge. In case i), $\Gamma_{WW}=0$ and the remaining widths have a fixed relative ratio,
\bea
\label{eq:case1}
\Gamma_{\gamma\gamma} : \Gamma_{Z\gamma}: \Gamma_{ZZ} &=& 1 : 2{s_w^2}/{c_w^2} : {s_w^4}/{c_w^4} \nonumber \\
&\approx&  1 :  0.6 : 0.09\ .
\eea
This particular  ratio  was observed by several groups previously \cite{Buttazzo:2015txu,Franceschini:2015kwy,Kobakhidze:2015ldh,Altmannshofer:2015xfo}.
Since the total width $\Gamma_{\rm tot} \ge \Gamma_{\gamma\gamma}+\Gamma_{Z\gamma}+\Gamma_{ZZ}$, there is an {\em upper bound} on the branching fraction into diphoton final states in this case
\be
{\rm BR}(S\to \gamma\gamma) \le 60 \% \ .
\ee
In case ii), all four channels are present and their relative ratio is also completely fixed
\bea
&&\Gamma_{\gamma\gamma} : \Gamma_{Z\gamma}: \Gamma_{ZZ}:\Gamma_{WW} \nonumber \\
&=& 1 : 2{c_w^2}/{s_w^2} : {c_w^4}/{s_w^4}: 2/s_w^4 \nonumber \\
&\approx& 1: 6.7: 11.2: 18.9 \ .
\eea
The upper limit on the diphoton decay BR is
\be
{\rm BR}(S\to \gamma\gamma) \le 2.6 \% \ .
\ee

However, experimentally it is easier to extract ratios of partial widths, which can be obtained from ratios of event rates
\be
\frac{B\sigma(gg\to S\to V_1V_2)}{B\sigma(gg\to S\to V^\prime_1V^\prime_2)} = \frac{\Gamma_{V_1V_2}}{\Gamma_{V^\prime_1V^\prime_2}}\ ,
\ee 
which only depends on the variable $\kappa_W/\kappa_B$. Therefore we can form the ratio of any two observed decay channels, which can be used to predict the ratios in the other two channels. This is demonstrated in the left panel of Fig.~\ref{fig:1}, where we normalize the partial widths in the $WW$, $ZZ$ and $Z\gamma$ channel to that of the diphoton channel and plot the resulting ratios as a function of $\kappa_W/\kappa_B$. Once a given ratio, say $\Gamma_{Z\gamma}/\Gamma_{\gamma\gamma}$ is measured, two possible values of $\kappa_W/\kappa_B$ are obtained, which in turn determine the other two ratios $\Gamma_{ZZ}/\Gamma_{\gamma\gamma}$ and $\Gamma_{WW}/\Gamma_{\gamma\gamma}$ that can be checked against experimentally. There are two possible values of $\kappa_W/\kappa_B$ for a given input, because the partial widths depend quadratically on $\kappa_B$ and $\kappa_W$.

%We can solve for $\kappa_W$ and $\kappa_B$ using $g_{S\gamma\gamma}$ and $g_{SZ\gamma}$:
%\bea
%\kappa_W&=& \frac{4\pi}{\alpha_{em}}s_wc_w\left(\frac{s_w}{c_w} g_{S\gamma\gamma} + g_{SZ\gamma}\right) \ , \\
%\kappa_B&=& \frac{4\pi}{\alpha_{em}}s_wc_w\left(\frac{c_w}{s_w} g_{S\gamma\gamma} - g_{SZ\gamma}\right) \ .
%\eea

It is interesting to note that one can derive an upper bound on the branching fraction to the diphoton channel for arbitrary $\kappa_B$ and $\kappa_W$.
Using Eq.~(\ref{eq:Gv1v2}) we can express the sub-total width for $S$ decays into electroweak gauge bosons as a function of $\kappa_W$ and $\kappa_B$:
\bea
&&\Gamma_{\rm ew}(\kappa_W,\kappa_B)\nonumber \\
& \equiv& \Gamma_{\gamma\gamma}+\Gamma_{Z\gamma}+\Gamma_{ZZ}+\Gamma_{WW} \nonumber \\
&=& \frac1{64\pi}\left(\frac{\alpha_{em}}{4\pi}\right)^2 m_S\left[a\, \kappa_B^2 +b\, \kappa_W^2 \right] \ ,
\eea
where
\bea
\label{eq:a}
a&=&1+2{s_w^2}/{c_w^2}+{s_w^4}/{c_w^4}\approx 1.67 \ ,\\
\label{eq:b}
b&=&1+ 2{c_w^2}/{s_w^2}+{(2+c_w^4)}/{s_w^4} \approx 56.71\ .
\eea
Then one can derive  an upper limit on the branching fraction of a singlet scalar decaying into diphoton for arbitrary $\kappa_W$ and $\kappa_B$:
\bea
\label{eq:diphob}
{\rm BR}(\gamma\gamma) =\frac{\Gamma_{\gamma\gamma}}{\Gamma_{\rm tot}} \le \frac{\Gamma_{\gamma\gamma}}{\Gamma_{\rm ew}}= \frac{(1+\kappa_W/\kappa_B)^2}{a+ b(\kappa_W/\kappa_B)^2 }\le   61\% \ ,
\eea
where the maximum occurs at 
\be
\label{eq:ratio}
\frac{\kappa_W}{\kappa_B} = \frac{a}{b}\approx 0.03\ ,
\ee
 which is very close to being saturated in case i). One can derive the upper bound on all other electroweak channels in a similar fashion.
 %Furthermore, after including other decay channels in the total width, the maximum decay branching fraction in the diphoton channel is still maximized given the above choice of $\kappa_W$ and $\kappa_B$. 
 In the right panel of Fig.~\ref{fig:1} we show the upper bound of ${\rm BR}(V_1V_2)$ as a function of $\kappa_W/\kappa_B$, which attains its maximum value of 61\% for $\kappa_W/\kappa_B\approx 0.03$ in the $\gamma\gamma$ channel and decreases quickly away from the maximum. In the $Z\gamma$, $ZZ$ and $WW$ channels, the maximum branching fractions are 47\%, 22\% and 67\%, respectively.
 
Fig.~\ref{fig:1} also exhibits two complementary decay patterns: near $\kappa_W\approx 0$, $\gamma\gamma$ and $Z\gamma$ channels are significantly enhanced over the $WW$ and $ZZ$ channels, which are the dominant ones away from $\kappa_W\approx 0$. Therefore, searches in the $WW$ and $ZZ$ channels probe complementary regions of parameter space from searches in $\gamma\gamma$ and $Z\gamma$ channels.

 %In the left panel we plot the partial widths in the $WW$, $ZZ$ and $Z\gamma$ channel in unit of $\Gamma_{\gamma\gamma}$. It can be seen that, in the region $\kappa_W/\kappa_B \approx 0$, the diphoton channel is the dominant mode among the four electroweak final states. However, as soon as $\kappa_W$ is turned on, $WW$ quickly becomes the dominant decay mode, followed by the $ZZ$ and $Z\gamma$ channel.

%%%%%%%%%%%%%%%%%%%%%%%%%%%%
\section{Implications on the scale of new particles}
%%%%%%%%%%%%%%%%%%%%%%%%%%%%%

%%%%%%%%%%%%%%%%%%%%%%%%%%%%%%%%%%%%%%%%%%%%%%%%%%
\begin{figure*}[!t]
  \begin{center}
    \includegraphics[width=0.47\textwidth]{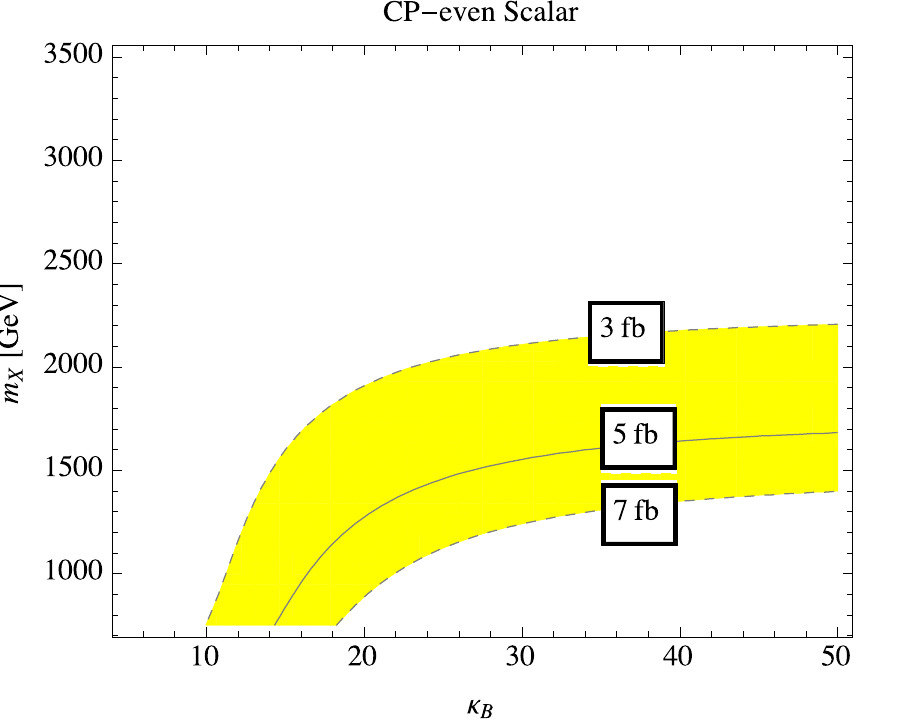}
        \includegraphics[width=0.47\textwidth]{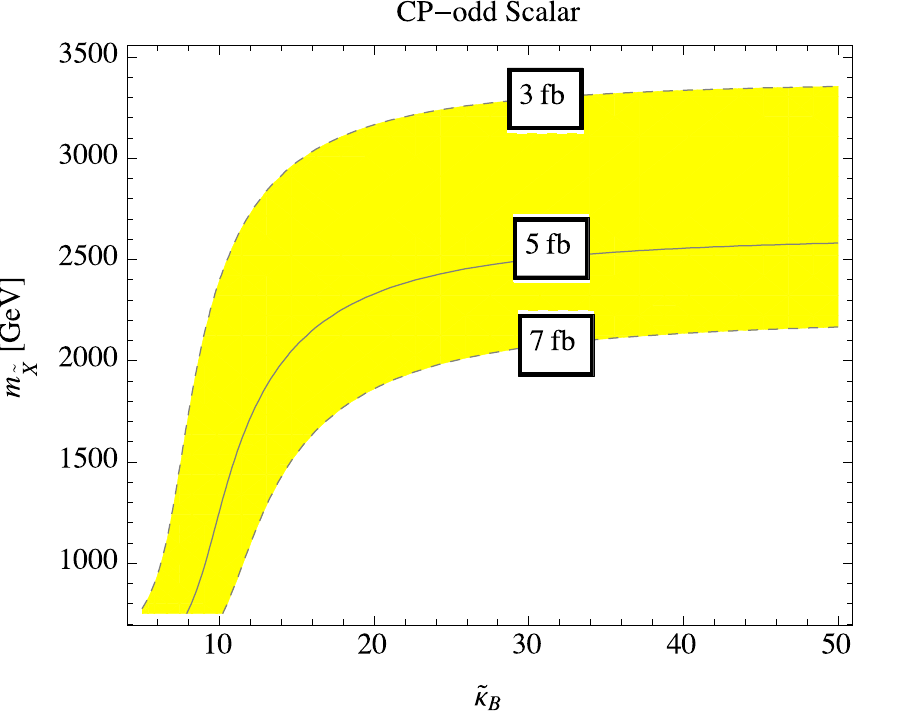}
    \caption{\label{fig:2} Upper bounds on the mass of the new colored scalar mediating the singlet scalar coupling to gluons, assuming an event rate $B\sigma(gg\to S/\tilde{S}\to \gamma\gamma)$ that is between 3 fb and 7 fb at 13 TeV LHC. Here a maximum diphoton decay branching fraction following from $\kappa_W/\kappa_B \approx 0.03 + 32.2 {\kappa_g^2}/{\kappa_B^2}$ is used.}
    \end{center}
\end{figure*}
%%%%%%%%%%%%%%%%%%%%%%%%%%%%%%%%%%%%%%%%%%%%%%%%%%

If we further assume that the diphoton resonance is produced predominantly in the gluon fusion channel, the upper bound on the diphoton fraction can be used to infer the mass scale of the new particle mediating the loop-induced couplings.  Given the experimental fit of the recent diphoton excess at 750 GeV:
\be
\sigma(gg\to S)\times {\rm BR}(S\to \gamma\gamma) \sim 5 \ {\rm fb} \ ,
\ee
a lower bound on the production cross-section is obtained, given the upper bound of the diphoton branching fraction, 
\be
\label{eq:crudeb}
\sigma(gg\to S) \agt 8.2 \ {\rm fb} \ .
\ee
This bound can be translated into an estimate on the mass scale of new colored particle mediating the gluon fusion production as follows. The coupling of the SM Higgs with gluons has the form
\be
\frac{\alpha_s}{12\pi}\frac{h}{v}G_{\mu\nu}^a G^{a\, \mu\nu} \ ,
\ee
and gives rise to a cross-section of $734$ fb at $m_h$=750 GeV. This number is obtained by multiplying the 8 TeV cross-section of 156.8 fb \cite{Heinemeyer:2013tqa} with the parton luminosity ratio of 4.7 \cite{parton}. Therefore, for a CP-even singlet scalar with the coupling to gluons in Eq.~(\ref{eq:su2u1inv}), the production cross-section is
\be
\sigma(gg\to S) = \left(\frac{3v}{4m_S}\right)^2 \kappa_g^2\times 734\ {\rm fb} =\kappa_g^2\times 44.4\ {\rm fb}\ .
\ee
The lower bound in Eq.~(\ref{eq:crudeb}) then requires $\kappa_g \agt 0.43$. Since, on general ground, $\kappa_g \sim m_S/m_X$, we then obtain
\be
\label{eq:mx}
m_X \alt 1.7\ {\rm TeV} \ .
\ee
For a CP-odd singlet scalar, if the colored particle mediating the gluonic coupling is a fermion, the loop function is a factor of 3/2 larger than the corresponding loop function for a CP-even scalar \cite{Djouadi:2005gj}. In this case the production cross-section is
\be
\sigma(gg\to \tilde{S}) = \left(\frac32 \frac{3v}{4m_{\tilde{S}}}\right)^2 \tilde{\kappa}_g^2\times 734\ {\rm fb}= \tilde{\kappa}_g^2 \times 99.9\ {\rm fb} \ .
\ee
Therefore the bound on the mass of the new colored particle is
\be
\label{eq:mxti}
m_{\tilde{X}} \alt 2.6 \ {\rm TeV} \ .
\ee
Notice that the lower bounds in Eqs.~(\ref{eq:mx}) and (\ref{eq:mxti}) are saturated when $\kappa_B \gg \kappa_g$, which requires either a large multiplicity of new charged particles that are uncolored, or a small number of new particles with substantial electric charges.

Away from the limiting case of $\kappa_B \gg \kappa_g$, the upper bound on the mass of new colored particle can be refined by including the gluonic decay width in the total width of the singlet scalar.
\be
\label{eq:maxBR}
{\rm BR}(\gamma\gamma) =\frac{(\kappa_W+\kappa_B)^2}{a \kappa_B^2+b\kappa_W^2+c\kappa_g^2} \ ,
%\le {\rm BR}(\gamma\gamma)\right|_{\frac{\kappa_W}{\kappa_B}=\frac{a}{b}} \ ,
\ee
where $a$ and $b$ are given in Eqs.~(\ref{eq:a}) and (\ref{eq:b}), respectively, and $c=8(\alpha_s/\alpha_{em})^2\approx 1827.2$. For a given $\kappa_g$, Eq.~(\ref{eq:maxBR}) is maximized for $\kappa_W/\kappa_B$ 
\be
\frac{\kappa_W}{\kappa_B} = \frac{a}{b} + \frac{c}{b} \frac{\kappa_g^2}{\kappa_B^2} \approx 0.03 + 32.2 \frac{\kappa_g^2}{\kappa_B^2} \ ,
\ee
from which we can compute the maximum possible event rate for a given $\kappa_g$, as a function of $\kappa_g$ and $\kappa_B$, 
\be
\left.\phantom{\frac{1}{2}}B\sigma \le \sigma(gg\to S)\times{\rm BR}(\gamma\gamma)\right|_{\frac{\kappa_W}{\kappa_B}=\frac{a}{b}+ \frac{c}{b} \frac{\kappa_g^2}{\kappa_B^2}} \ .
\ee
In Fig.~\ref{fig:2} we plot the right-hand side of the above equation with respect to $\kappa_B$ and $m_X \sim m_S/\kappa_g$, for both the CP-even and the CP-odd scalars. We see that the additional colored scalars need to be lighter than 2.7 TeV (3.4 TeV) for a CP-even (CP-odd) scalar, if we assume a smaller event rate of about 3 fb for the 750 GeV diphoton excess. If we take the event rate to be about 5 fb, the bounds become 1.7 TeV and 2.6 TeV, respectively.

\section{Discussion and outlook}
\label{sect:section5}

In this note we have focused on the possibility that the ATLAS and CMS diphoton excesses are the result of a heavy electroweak singlet scalar that couples to SM gauge bosons via dimension five operators gauge invariant under $SU(2)_L\times U(1)_Y$. We considered a CP-even scalar, but showed how similar arguments apply to a CP-odd scalar. For a CP-even scalar we made the non-trivial assumption that we can ignore mixing with the SM Higgs. In other respects our treatment was model independent.

In our simple framework experimental measurements on any two of the four possible electroweak channels would determine the remaining two decay channels completely. Searches in the $WW/ZZ$ channels probe a complimentary region of parameter space from searches in the $\gamma\gamma/Z\gamma$ channels. We derived a model-independent upper bound on the branching fraction in each decay channel, which for the diphoton channel turns out to be about 61\%. We then made the further assumption that the singlet resonance is produced predominantly via gluon fusion, and showed that the upper bound on the diphoton branching fraction implies an upper bound on the mass scale of additional colored particles mediating the gluon-fusion production. Using an event rate of 5 fb for the reported diphoton excess, we found that the new colored particle must be lighter than ${\cal O}(1.7\ {\rm TeV})$ and ${\cal O}(2.6\ {\rm TeV})$ for a pure CP-even and a pure CP-odd singlet scalar, respectively; Fig.~\ref{fig:2} shows the scaling of these bounds when one makes other assumptions about the event rate.

It is of course premature to attempt to squeeze too much out of what may well turn out to be a pair of statistical fluctuations. But by the same token if a new heavy resonance is indeed present, new data could very quickly reveal much about its nature. Of particular interest are stronger constraints or high mass excesses in the other diboson channels $ZZ$, $WW$, $Z\gamma$; as observed in \cite{Altmannshofer:2015xfo}, 
analyses from LHC Run 1 \cite{Khachatryan:2015cwa,Aad:2015kna,Aad:2014fha,Aad:2015agg} already provide interesting constraints on the heavy singlet scenario discussed here.

\begin {acknowledgements}
I.~L. was supported in part by the U.S. Department of Energy under
contracts No. DE-AC02-06CH11357 and No. DE-SC 0010143.
Fermilab is operated by the Fermi
Research Alliance LLC under contract DE-AC02-07CH11359 with the
U.S. Department of Energy.
\end{acknowledgements}

\end{document}